\documentclass[useAMS,usenatbib]{mn2e}
\usepackage{aas_macros}
\usepackage{amsmath,amssymb}
\usepackage{bm}
\usepackage{color}
\usepackage[dvipdfmx]{hyperref}
\usepackage{graphicx}
\usepackage{enumerate}

\newcommand{\nnmb}{\nonumber\\}

\newcommand{\ds}{\displaystyle}

\begin{document}
\author[K. Sugimura, K. Omukai and A. K. Inoue]{Kazuyuki
Sugimura,$^1$\thanks{E-mail: sugimura@astr.tohoku.ac.jp} Kazuyuki
Omukai$^1$ and Akio K. Inoue$^2$ \\ $^1$Astronomical Institute, Tohoku
University, Aoba, Sendai 980-8578, Japan \\ $^2$College of General
Education, Osaka Sangyo University, Daito, Osaka
574-8530, Japan} 
\title[$J^{\rm crit}$ for direct collapse BH
formation]{The critical radiation intensity for direct collapse black
hole formation: dependence on the radiation spectral shape
}
\maketitle

\begin{abstract}
It has been proposed that supermassive black holes (SMBHs) are
originated from direct-collapse black holes (DCBHs) that are formed at
$z\gtrsim 10$ in the primordial gas in the case that $\rm H_2$ cooling
is suppressed by strong external radiation.  In this work, we study the
critical specific intensity $J^{\rm crit}$ required for DCBH formation
for various radiation spectral shapes by a series of one-zone
calculations of a collapsing primordial-gas cloud.  We calculate the
critical specific intensity at the Lyman-Werner (LW) bands $J_{\rm
LW,\,21}^{\rm crit}$ (in units of $10^{-21}\, {\rm erg}\, {\rm s}^{-1}\,
{\rm Hz}^{-1}\, {\rm sr}^{-1}\,{\rm cm}^{-2}$) for realistic spectra of
metal-poor galaxies.  We find $J^{\rm crit}$ is not sensitive to the age
or metallicity for the constant star formation galaxies with $J_{\rm
LW,\,21}^{\rm crit}=1300-1400$, while $J^{\rm crit}$ decreases as
galaxies become older or more metal-enriched for the instantaneous
starburst galaxies.  However, such dependence for the instantaneous
starburst galaxies is weak for the young or extremely metal-poor
galaxies: $J_{\rm LW,\,21}^{\rm crit}= 1000-1400$ for the young (the age
less than 100 Myr) galaxies and $J_{\rm LW,\,21}^{\rm crit}\approx 1400$
for the extremely metal-poor ($Z<5\times 10^{-4}Z_\odot$) galaxies.  We
also find $J^{\rm crit}$ is solely determined by the ratio of the $\rm
H^-$ and ${\rm H_2}$ photodissociation rate coefficients $k_{\rm
H^-,\,pd}/k_{\rm H_2,\,pd}$, with which we develop a formula to estimate
$J^{\rm crit}$ for a given spectrum.  The typical value of $J^{\rm
crit}$ for the realistic spectra is higher than those expected in the
literature, which affects the estimated DCBH number density $n_{\rm
DCBH}$.  By extrapolating the result of Dijkstra, Ferrara \& Mesinger,
we obtain $n_{\rm DCBH}\sim 10^{-10}{\rm cMpc^{-3}}$ at $z=10$, although
there is still large uncertainty in this estimation.  This estimated
$n_{\rm DCBH}$ is roughly consistent with the observed number density of
high-redshift SMBHs $n_{\rm SMBH} \sim 10^{-9} {\rm cMpc^{-3}}$ at
$z\sim 6$, considering uncertainties, but much less than that of
present-day SMBHs $n_{\rm SMBH}\sim 10^{-4}{\rm cMpc^{-3}}$, indicating
other seed BH formation mechanisms are also operating.

\end{abstract}


\section{Introduction}
\label{sec:introduction}

Observations reveal that almost all galaxies host supermassive black
holes (SMBHs) at their centers
(\cite{Ferrarese:2000aa,Gebhardt:2000aa,Gultekin:2009aa}).  Although
SMBHs play an important role in the cosmic history by their radiative
activities via accretion of surrounding gas, their origin is
remained one of the most puzzling mysteries in astrophysics.  The
discovery of SMBHs with inferred black hole mass $> 10^9M_\odot$ at $z>6$
(\cite{Fan:2001aa,Mortlock:2012aa,Venemans:2013aa}) suggests that SMBH
seeds are formed very early in the history of the Universe.  In order
for remnants of first generation (pop III) stars ($M_{\rm popIII}\sim
100M_\odot$) to be seeds of SMBHs, surrounding gas is needed to accrete
at the Eddington limited rate for the entire period of accretion from
$\sim100M_\odot$ to $>10^9M_\odot$. However, the Eddington limited accretion
is likely to be prevented by radiative feedback 
(\cite{Johnson:2007aa,Alvarez:2009aa,Milosavljevic:2009aa}).

A possible solution to this problem is that SMBH seeds are not remnants
of pop III stars but direct collapse black holes (DCBHs) that are formed
by direct collapse of supermassive stars (SMSs) with mass
$\gtrsim10^5M_\odot$ (\cite{Bromm:2003aa}).  SMSs are expected to be
formed from primordial-gas clouds in halos with virial temperature $T_{\rm vir}
\gtrsim 10^4\,{\rm K}$, in the case that the clouds collapse 
isothermally with the temperature of gas $T_{\rm gas}\sim 8000\,{\rm K}$
via atomic cooling in the
absence of $\rm H_2$ molecules due to strong external radiation
(\cite{Omukai:2001aa} hereafter O01).\footnote{Other physical mechanism
such as shock heating of gas
(\cite{Inayoshi:2012aa,Visbal:2014ac}) are proposed to
suppress $\rm H_2$ formation, but we concentrate on the case of strong
external radiation in this paper.} In such a case, it has been shown that
fragmentation of the gas is suppressed
(\cite{Bromm:2003aa,Regan:2007aa,Regan:2009aa,Inayoshi:2014aa}) and that
the large accretion rate
is expected to continue until SMSs (and subsequently DCBHs) are formed
(\cite{Hosokawa:2012aa,Hosokawa:2013aa}).

In this paper, we calculate the critical specific intensities of external 
radiation $J^{\rm crit}$ required for DCBH formation.  
External radiation reduces the abundance of $\rm H_2$ in two ways: one
is by direct photodissociation of $\rm H_2$ with the Lyman-Werner (LW) photons
(photons with energies $11.2\,{\rm eV} < h\nu < 13.6\,{\rm eV}$); the other
is by photodissociation of the intermediary $\rm H^-$ of the dominant $\rm
H_2$ formation channel with the photons with energies $\gtrsim 0.76\,{\rm
eV}$.  For the fixed spectral shape of radiation, $J^{\rm crit}$ is defined 
as the critical specific intensity $J_{\rm LW,\,21}^{\rm
crit}\equiv J_{21}^{\rm crit}(h\nu= 12.4\,{\rm eV})$ (in units of $10^{-21}
\,{\rm erg}\, {\rm s}^{-1}\, {\rm Hz}^{-1}\, {\rm sr}^{-1}\,{\rm
cm}^{-2}$) at the center of the LW bands ($11.2\,{\rm eV} < h\nu <
13.6\,{\rm eV}$).\footnote{Some literature defined the critical specific
intensity as $J_{\rm Lyc,\,21}^{\rm crit}\equiv J_{21}^{\rm
crit}(h\nu=13.6\,{\rm eV})$ at the limit of the Lyman continuum
(Lyc), but small difference in two definitions doesn't matter in the
order-of-magnitude argument in the introduction.}  The critical intensity $J^{\rm crit}$ has
been obtained in the various physical conditions, by using one-zone calculations (O01;
\cite{Omukai:2008aa,Inayoshi:2011aa,Wolcott-Green:2011ab}) or
three-dimensional hydrodynamic simulations (\cite{Shang:2010aa} hereafter S10; \cite{Latif:2014ab}).

The feasibility of the SMBH formation scenario via DCBH can be tested by
comparing the estimated DCBH number density $n_{\rm DCBH}$ with the
observed high-redshift SMBH number density $n_{\rm SMBH} \sim 10^{-9}
{\rm cMpc^{-3}}$ at $z\sim 6$ (\cite{Fan:2001aa,Venemans:2013aa}).
Although $J^{\rm crit}$ has wide range of varieties depending on
physical conditions (O01; \cite{Omukai:2008aa}; S10;
\cite{Inayoshi:2011aa,Wolcott-Green:2011ab,Latif:2014ab}), they are in
general much higher than the averaged cosmic LW background in the whole
history of the Universe (see, e.g., \cite{OShea:2008aa,Johnson:2013ab}).
To be concrete, $J_{\rm LW,\,21}^{\rm crit}=O(10)-O(10^4)$ while
$J_{\rm bg,\,LW,\,21}\lesssim 0.1$.  Thus, $J>J^{\rm crit}$ is
achievable only in the rare situations that a primordial-gas cloud is
irradiated by strong radiation from unusually nearby and/or bright
galaxies, and the fraction  of primordial-gas clouds
with $T_{\rm vir}\gtrsim 10^4\,{\rm K}$ that can form DCBHs $f(J >J^{\rm crit})$ is very small.
In the literature (\cite{Dijkstra:2008aa}; \cite{Agarwal:2012aa}
hereafter A12; \cite{Agarwal:2014aa}; \cite{Dijkstra:2014aa} hereafter
D14; \cite{Yue:2014aa}), $f(J >J^{\rm crit})$ was obtained from semi-analytical calculations
to estimate $n_{\rm DCBH}$.
They showed that the clouds with $J >J^{\rm crit}$ are distributed at
the high $J$ tail of the probability density, and that even a small
change in the value of $J^{\rm crit}$ causes significant difference to
the predicted value of $n_{\rm DCBH}$. Thus, precise determination of
$J^{\rm crit}$ is very important in estimating $n_{\rm DCBH}$.

It is known that $J^{\rm crit}$ strongly depends on the
spectral shape of external radiation(O01).  While $J_{\rm
LW,\,21}^{\rm crit}=O(10)$ for the black-body spectrum with $T_{\rm
rad}=10^4\,{\rm K}$ (S10), $J_{\rm LW,\,21}^{\rm crit}=O(1000)$ for that
with $T_{\rm rad}=10^5\,{\rm K}$ (\cite{Wolcott-Green:2011aa} hereafter
WG11). The black-body spectra with $T_{\rm rad}=10^4\,{\rm K}$ and $10^5{\rm
K}$ have frequently been used as approximate spectra of Pop II and Pop III galaxies, respectively, 
in the literature. 
However, the hardness of realistic spectra ranges between
that of the above two black-body spectra
(\cite{Leitherer:1999aa,Schaerer:2003aa,Inoue:2011aa}), and thus actual
values of $J^{\rm crit}$ realized in the Universe are not clear yet.  In
this paper, we study the dependence of $J^{\rm crit}$ on spectra
and obtain $J^{\rm crit}$ for realistic spectra of galaxies calculated
by the stellar population synthesis models
(\cite{Schaerer:2003aa,Leitherer:1999aa,Inoue:2011aa}).

This paper is organized as follows.  In Sec.~\ref{sec:model}, we
describe our one-zone model used to calculate the evolution of
primordial-gas clouds under external radiation.  In
Sec.~\ref{sec:evolution}, we review physical processes proceeding during
the evolution of the clouds, showing several results of our one-zone calculations.  We determine $J^{\rm crit}$ for the black-body spectra with
various temperatures in Sec.~\ref{sec:t-dependence}, and for realistic
spectra in Sec.~\ref{sec:realistic}.  In Sec.~\ref{sec:fitting-formula},
we find the key parameter determining the dependence of $J^{\rm crit}$
on spectra, and develop a formula to estimate $J^{\rm crit}$ based on
this parameter.  Finally, we present the summary and discussion of this
work in Sec.~\ref{sec:summary-discussion}.

\section{Model}
\label{sec:model}

\subsection{Basics}
\label{sec:basics}
In this paper, we use a one-zone model, as described in O01, to follow
the gravitational collapse of primordial-gas clouds.  By neglecting
effects due to rotation or magnetic fields for simplicity, the
gravitational collapse is expected to proceed like the self-similarity
solution (\cite{Penston:1969aa,Larson:1969aa,Yahil:1983aa}).  It has
been confirmed that this simplified dynamical evolution actually
describes the essential part of the gravitational collapse in
three-dimensional hydrodynamic simulations (S10; \cite{Latif:2014ab}).
The quantities computed in one-zone models correspond to those in the
nearly homogeneous central core of the self-similarity solution.  The
chemical, thermal and radiative processes are solved in detail.  In the
followings, we briefly explain the basics of our one-zone model, which is
almost the same as the literature (O01; \cite{Omukai:2008aa}; S10), 
but with updated microphysics.

For the dynamical evolution, we assume the collapse of clouds proceeds as
\begin{align}
 \frac{d\rho_{\rm B}}{dt} = \frac{\rho_{\rm B}}{t_{\rm ff}}\,,
\end{align}
where $t_{\rm ff} \equiv \sqrt{3\pi/32 G\rho}$ is the free-fall time, $G$ the gravitational constant, $\rho=\rho_{\rm B}+\rho_{\rm DM}$ the total density,
$\rho_{\rm B}$ the baryonic density and $\rho_{\rm DM}$ the dark matter (DM) density.
 We assume the evolution of $\rho_{\rm
DM}$ is described by the spherical top-hat collapse model until $\rho_{\rm DM}$
reaches the virial density (see, e.g., O01). After that we keep $\rho_{\rm DM}$ constant.
We assume the size of central core equals the Jeans length,
\begin{align}
\lambda_{\rm J}=\sqrt{\frac{\pi k T_{\rm gas} } { G \rho_{\rm B}\mu m_{\rm H}}}\,,
\label{eq:21}
\end{align}
where $m_{\rm H}$ is the proton mass, $\mu$ the mean molecular weight and
$k$ the Boltzmann constant.

For the chemical evolution in primordial-gas clouds, we solve the
chemical network of 9 species, ${\rm H}$, ${\rm H_2}$, ${\rm e}$, ${\rm
H}^+$, ${\rm H}^-$, ${\rm H}_2^+$, ${\rm He}$, ${\rm He}^+$ and ${\rm
He}^{++}$, with the chemical reaction rates of \cite{Glover:2008aa}.  In
this work, we do not consider deuterium since the inclusion of it should
make no difference to our results.  Deuterium becomes important only in
the situation that the gas is cooled below a few hundred Kelvin by HD
cooling (see, e.g., \cite{Nagakura:2005aa,McGreer:2008aa,Nakauchi:2014aa}).  In the followings, we denote
the number density of hydrogen nuclei as $n$, that of helium nuclei as
$n_{\rm He}$ and that of species $A$ as $n(A)$. We also denote the
abundance of species $A$ normalized by $n$ as $y(A)\equiv n(A)/n$.  The
chemical evolution is affected by external radiation via
photodissociation processes, which we explain in detail in
Sec.~\ref{sec:UV_rad}.

The temperature evolution is described by the energy equation,
\begin{align}
 \frac{de}{dt}= -p \frac{d}{dt}\left(\frac{1}{\rho_{\rm B}}\right)
 - \frac{\Lambda_{\rm net}}{\rho_{\rm B}}\,,
\end{align}
where $e= p/\rho_{\rm B}(\gamma_{\rm ad}-1)$ is the internal energy per
unit mass of baryon, $p=\rho_{\rm B}kT_{\rm gas}/\mu m_{\rm H}$ the
pressure and $\gamma_{\rm ad}$ the adiabatic exponent.
The net cooling rate per unit volume $\Lambda_{\rm net}$ is given
by $\Lambda_{\rm net}=\Lambda_{\rm H}+\Lambda_{\rm H_2}+\Lambda_{\rm
chem}$, where $\Lambda_{\rm H}$, $\Lambda_{\rm H_2}$ and $\Lambda_{\rm
chem}$ are the cooling rates due to radiative cooling by Ly$\alpha$
(\cite{Anninos:1997aa}) and ${\rm H_2}$
(\cite{Glover:2008aa} with the LTE value by \cite{Hollenbach:1979aa}) and  due to
chemical reaction (\cite{Shapiro:1987aa}), respectively.

We start the calculation at the turnaround time, when the motion of the
gas and DM turns from expansion to collapse.  We assume that the turnaround
time is at $z=16$ and that initial values for physical quantities are
given by $n=4.5\times 10^{-3} {\rm cm}^{-3}$, $T_{\rm gas}=21\,{\rm K}$,
the ionizing degree $y({\rm e})=3.7\times10^{-4}$ and the ${\rm H_2}$ fraction
$y({\rm H_2})= 2\times 10^{-6}$, reflecting the condition of the
universe at $z=16$ (\cite{Omukai:2008aa}).  It has been confirmed that
the results are almost independent of the initial conditions as long as realistic 
values are chosen (\cite{Omukai:2008aa}).

\subsection{The effects of external radiation on the ${\rm H_2}$ abundance}
\label{sec:UV_rad}
In this section, we briefly review the key processes determining the
${\rm H_2}$ abundance under the influence of external radiation (for
more detailed review, see e.g. O01).  As explained in the introduction,
primordial-gas clouds collapse via atomic cooling in the case that strong
external radiation suppresses $\rm H_2$ cooling. In the followings, we
review two ${\rm H_2}$ formation and dissociation channels and three
photodissociation processes.

Let us start with reviewing the ${\rm H_2}$ formation channel via intermediary ${\rm H^-}$, 
which is the dominant ${\rm H_2}$ formation channel in most cases.
This channel begins with the ${\rm H^-}$ formation reaction, 
\begin{align}
 {\rm H} + e \rightarrow {\rm H^-} + {\rm \gamma}\,,
 \label{eq:7}
\end{align}
which is followed by the ${\rm H_2}$ formation reaction,
\begin{align}
 {\rm H^-} + {\rm H} \rightarrow  {\rm H_2} +  e \,.
 \label{eq:8}
\end{align}
We denote the reaction rate coefficients for Eqs.~\eqref{eq:7} and \eqref{eq:8}
as $k_{\rm form}^{(1)}\,{\rm [cm^3 s^{-1}]}$ and $k_{\rm form}^{(2)}\,{\rm [cm^3 s^{-1}]}$, respectively.
In the above chain, not all ${\rm H^-}$ molecules formed via Eq.~\eqref{eq:7} are used for
${\rm H_2}$ formation but some of them go back to H by
the ${\rm H^-}$ photodissociation reaction,
\begin{align}
 {\rm H^-} + {\rm \gamma} \rightarrow {\rm H} + e\,.
 \label{eq:11}
\end{align}
Here, the photodissociation rate coefficient, denoted as $k_{\rm H^-,\,pd}\,[{\rm s}^{-1}]$, is proportional to the number density of photons
of external radiation. The rates of competing reactions given by
Eqs.~\eqref{eq:8} and \eqref{eq:11} determine the branching ratio of
formed ${\rm H^-}$ to be used for ${\rm H_2}$ formation.  Since the
reactions of Eqs.~\eqref{eq:8} and \eqref{eq:11}
proceeds much faster than that of
Eq.~\eqref{eq:7}, the formation rate of ${\rm H_2}$ per unit volume per
unit time can be written as $ k_{\rm form}^{\rm (eff)}\,n({\rm
H})\,n({\rm e})$, where the effective ${\rm H_2}$ formation rate
coefficient $k_{\rm form}^{\rm (eff)}\,{\rm [cm^3 s^{-1}]}$ is given by
\begin{align}
 k_{\rm form}^{\rm (eff)}\equiv
 k_{\rm form}^{(1)} \left[\frac{k_{\rm form}^{(2)}\, n({\rm H})}{k_{\rm form}^{(2)}\, n({\rm H})+k_{\rm H^-,\,pd}}\right]\,.
 \label{eq:10}
\end{align}

Next, we would like to review another ${\rm H_2}$ formation channel via
intermediary ${\rm H_2^+}$, which is less effective than the ${\rm H_2}$
formation channel via ${\rm H^-}$ in most cases.  
This channel begins with the ${\rm H_2^+}$ formation reaction, 
\begin{align}
 {\rm H} + {\rm H^+} \rightarrow {\rm H_2^+} + {\rm \gamma}\,,
\label{eq:16}
\end{align}
which is followed by the ${\rm H_2}$ formation reaction,
\begin{align}
 {\rm H_2^+} + {\rm H} \rightarrow  {\rm H_2} +   {\rm H^+}\,.
\label{eq:18}
\end{align}
This channel can be regarded as an analogue of the ${\rm H_2}$ formation
channel via ${\rm H^-}$. In this case, however, the reaction chain
begins with collision of ${\rm H}$ with ${\rm H^+}$
instead of $e$. In a similar way to the ${\rm H_2}$ formation
channel via ${\rm H^-}$, not all the ${\rm H_2^+}$ molecules
formed via Eq.~\eqref{eq:16} are used for
${\rm H_2}$ formation due to the ${\rm H_2^+}$ photodissociation reaction,
\begin{align}
{\rm H_2^+} + {\rm \gamma} \rightarrow  {\rm H} + {\rm H^+}\,,
\label{eq:20}
\end{align}
where the photodissociation rate coefficient $k_{\rm H_2^+,\,pd}\,[{\rm s}^{-1}]$ is
proportional to the density of photons of external radiation.  Here, again, the rates
of competing reactions given by Eqs.~\eqref{eq:18} and \eqref{eq:20}
determine the branching ratio of formed ${\rm H_2^+}$ to be used for
${\rm H_2}$ formation.  In principle, the $\rm H_2$ formation channel
via ${\rm H_2^+}$ can overwhelm that via ${\rm H^-}$ by suppressing only
latter by ${\rm H^-}$ photodissociation.  However, it is unlikely to be
realized in our calculations since the strength of ${\rm H_2^+}$  and
${\rm H^-}$ photodissociation are closely related, as explained in the
last part of this section.

The main ${\rm H_2}$ dissociation channel changes depending on the density of gas $n$.
When $n$ is small, the dominant channel is the ${\rm H_2}$ photodissociation reaction,
\begin{align}
 {\rm H_2} + {\rm \gamma} \rightarrow {\rm H_2^*}\rightarrow 2{\rm H}\,,
 \label{eq:14}
\end{align}
where the ${\rm H_2}$ photodissociation rate coefficient
$k_{\rm H_2,\,pd}[{\rm s}^{-1}]$ is proportional to the density of photons of external radiation.
On the other hand, when $n$ is large, the dominant channel 
is the collisional dissociation reaction,
\begin{align}
 {\rm H_2} + {\rm H} \rightarrow 3{\rm H}\,,
 \label{eq:13}
\end{align}
where we denote the collisional dissociation rate coefficient as $k_{\rm cd,\, H_2}{\rm [cm^3 s^{-1}]}$.

In the followings, we review the three photodissociation processes due to
external radiation: ${\rm H_2}$ , ${\rm H^-}$  and $\rm
H_2^+$ photodissociation. 

\begin{figure}
\hspace*{-0.3cm} \includegraphics[width=9cm]{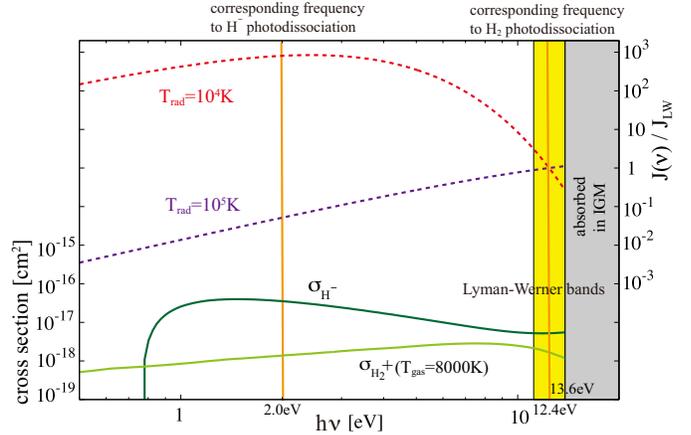}
 \caption{The cross sections for ${\rm H^-}$ photodissociation $\sigma_{\rm H^-}$ (green)
 and for ${\rm H_2^+}$ photodissociation when $T_{\rm gas}=8000\,{\rm K}$
 $\sigma_{\rm H_2^+}(T_{\rm
gas}=8000\,{\rm K})$ (light-green) are plotted. The LW bands, corresponding to ${\rm
 H_2}$ photodissociation, are drawn as an yellow band.  The black-body spectra
 $J(\nu)$ with $T_{\rm rad}=10^4\,{\rm K}$ and $10^5\,{\rm K}$ are also
 plotted in the same figure.  In this paper, we assume ionizing photons ($h\nu>13.6{\rm
 eV}$) are totally absorbed in the intergalactic
medium (IGM).  The specific intensities at $h\nu=12.4\,{\rm eV}$ and
 $2.0\,{\rm eV}$ are related to ${\rm H_2}$ and ${\rm H^-}$ photodissociation, respectively
 (see Eqs.~\eqref{eq:2} and \eqref{eq:22}, respectively).  The form of $\sigma_{\rm
 H^-}(\nu)$ given by \protect\cite{John:1988aa} can be used only for
 $h\nu<9.8\,{\rm eV}$, but we keep using it even for $h\nu>9.8\,{\rm eV}$
 since the error introduced by this treatment is expected to be negligibly small. }
 \label{fig:nu_sigma}
\end{figure}

First, let us review ${\rm
H_2}$ photodissociation given by Eq.~\eqref{eq:14}.
${\rm H_2}$ photodissociation is one of the key processes in our calculations
because it suppresses the ${\rm H_2}$ abundance by directly dissociating ${\rm H_2}$ molecules.
The photodissociation rate coefficient $k_{\rm H_2,\,pd}$ can be calculated
from external radiation $J(\nu)$ as
(\cite{Draine:1996aa})
\begin{align}
 k_{\rm H_2,\,pd}&\approx \kappa_{\rm H_2,\,pd}\,J_{\rm LW}\,,
 \label{eq:2}
\end{align}
with $\kappa_{\rm H_2,\,pd}=1.4\times10^9\,  {\rm (in\ cgs\ unit)}$.
Here, $k_{\rm H_2,\,pd}$ is estimated with $J_{\rm LW}\equiv
J(h\nu=12.4\,{\rm eV})$, the specific intensity at the center of LW bands
($11.2\,{\rm eV} < h\nu < 13.6\,{\rm eV}$), as shown in
Fig.~\ref{fig:nu_sigma}.  The error due to estimating $k_{\rm H_2,\,pd}$ by using the specific intensity at one frequency is usually negligible
since $J(\nu)$ does not change significantly in the narrow frequency
range of the LW bands.

The intensity in the LW bands is self-shielded by ${\rm H_2}$ molecules
when the ${\rm H_2}$ column density of the central core $N_{\rm H_2}$
becomes large.  We take this effect into account in our one-zone model
by multiplying the intensity in the LW bands by a self-shielding factor
$f_{\rm sh}$. It seems that there is no complete agreement on the form
of $f_{\rm sh}$ yet (see WG11 and \cite{Richings:2014aa} hereafter R14),
although, in principle, it should be determined uniquely by studying the
effective amount of self-shielding with level-by-level radiative
transfer calculations.  Considering such situation, we decide to use the
form of $f_{\rm sh}$ derived in WG11 as a fiducial model, but to study
the influence of using different forms in Sec.~\ref{sec:fsh-dependence}.
The form of $f_{\rm sh}$ derived in WG11 is
\begin{align}
 f_{\rm sh}(N_{\rm H_2}, T_{\rm gas})&=
\frac{0.965}{(1+x/b_5)^{1.1}}+\frac{0.035}{(1+x)^{0.5}}\nnmb
&\hspace{0.5cm}\times \exp\left[-8.5\times10^{-4}(1+x)^{0.5}\right]\,.
\label{eq:12}
\end{align}
where
\begin{align}
x&\equiv \frac{N_{\rm H_2}}{5\times 10^{14}{\rm cm^{-2}}}\,,\qquad
b_5\equiv     \frac{\sqrt{2kT_{\rm gas}/2m_{\rm H}}}{10^5{\rm cm\, s^{-1}}}\,.
\label{eq:13}
\end{align}
We assume $N_{\rm H_2}$ is given by
\begin{align}
N_{\rm H_2}=n({\rm H_2})\,\lambda_{\rm J}\,,
\end{align}
with the Jeans length $\lambda_{\rm J}$ given by Eq.~\eqref{eq:21}.

Second, let us review ${\rm H^-}$ photodissociation given by 
Eq.~\eqref{eq:11}.
${\rm H^-}$ photodissociation is also one of the key processes in our calculations
because it suppresses the dominant ${\rm H_2}$ formation channel by 
dissociating intermediary ${\rm H^-}$.
The photodissociation rate coefficient $k_{\rm H^-,\,pd}$ is calculated from $J(\nu)$ as
\begin{align}
 k_{\rm H^-,\,pd}=\int_0^\infty \frac{4\pi J(\nu)}{h\nu}\sigma_{\rm H^-}(\nu)d\nu\,,
 \label{eq:3}
\end{align}
with the cross section $\sigma_{\rm H^-}(\nu)$ of \cite{John:1988aa}.
We need to evaluate the frequency integral in Eq.~\eqref{eq:3} to obtain
$k_{\rm H^-,\,pd}$, since ${\rm H^-}$ photodissociation is caused by a
wide range of photons ($0.76\,{\rm eV} < h\nu $), as shown in
Fig.~\ref{fig:nu_sigma}.  For later convenience, however, we introduce a
similar expression to Eq.~\eqref{eq:2},
\begin{align}
 k_{\rm H^-,\,pd}&\approx  \kappa_{\rm H^-,\,pd}\,  J_{\rm 2eV}\,,
\label{eq:22}
\end{align}
where $J_{\rm 2eV}\equiv J(h\nu=2.0\,{\rm eV})$ and the information of the
spectral shape of external radiation is contained in $\kappa_{\rm H^-,\,pd}$.
We note that $h\nu=2.0\,{\rm eV}$ is the
frequency above and below which the integration in Eq.~\eqref{eq:3}
is equal for a flat spectrum ($J(\nu)={\rm const.}$).

Finally, let us comment on ${\rm H_2^+}$ photodissociation given by
Eq.~\eqref{eq:20}.  The $\rm H_2^+$ channel of the $\rm H_2$ formation
may become the dominant channel when $\rm H^-$ photodissociation
suppresses the $\rm H^-$ channel, which is dominant without any
photodissociation.  In the followings, we check which channel is
dominant, in the case that the $\rm H_2^+$ channel, as well as the $\rm
H^-$ channel, is suppressed by photodissociation.  We use the cross
section $\sigma_{\rm H_2^+}(\nu,T_{\rm gas})$ given by
\cite{Stancil:1994aa} ($T_{\rm gas}>2000\,{\rm K}$) and
\cite{Mihajlov:2007aa} ($T_{\rm gas}<2000\,{\rm K}$). The cross section
$\sigma_{\rm H_2^+}$ depends on $T_{\rm gas}$ because ${\rm H_2^+}$ is
easier to be dissociated from excited states, which are assumed to be
populated according to the LTE distribution with $T_{\rm gas}$.  In the
case that $T_{\rm gas}\sim 8000\,{\rm K}$, the frequency range
contributing to ${\rm H_2^+}$ photodissociation is wider than that
contributing to ${\rm H^-}$ photodissociation, while $\sigma_{\rm
H_2^+}$ is smaller than $\sigma_{\rm H^-}$ by an order of magnitude at
$h\nu>0.76\,{\rm eV}$, as shown in Fig.~\ref{fig:nu_sigma}, and thus the
frequency integrated $\rm H_2^+$ photodissociation rate coefficient
$k_{\rm H_2^+,\,pd}$, defined in a similar way to Eq.~\eqref{eq:3},
becomes also large in the case $k_{\rm H^-,\,pd}$ is large. Therefore,
the $\rm H_2^+$ channel is always subdominant even if the
photodissociation processes are considered.

\section{Results}
\label{sec:results}

\subsection{The evolution of primordial-gas clouds under external radiation}
\label{sec:evolution}

In this section, we review physical processes proceeding during the
gravitational collapse of primordial-gas clouds under external radiation,
showing several results of our one-zone calculations.  Following the
argument in O01, we see how the evolutionary trajectories bifurcate to the
atomic  and $\rm H_2$ cooling tracks depending on the strength and spectral shape
of external radiation.  We make calculations of
collapsing clouds irradiated by the black-body spectra with $T_{\rm
rad}=10^4\,{\rm K}$ and $T_{\rm rad}=10^5\,{\rm K}$.  We specify
the strength of radiation using $J_{\rm LW,\,21}$, in units of $10^{-21} {\rm
erg}\, {\rm s}^{-1}\, {\rm Hz}^{-1}\, {\rm sr}^{-1}\,{\rm cm}^{-2}$.

\begin{figure}
 \centering \hspace*{-0.3cm}
\includegraphics[width=9cm]{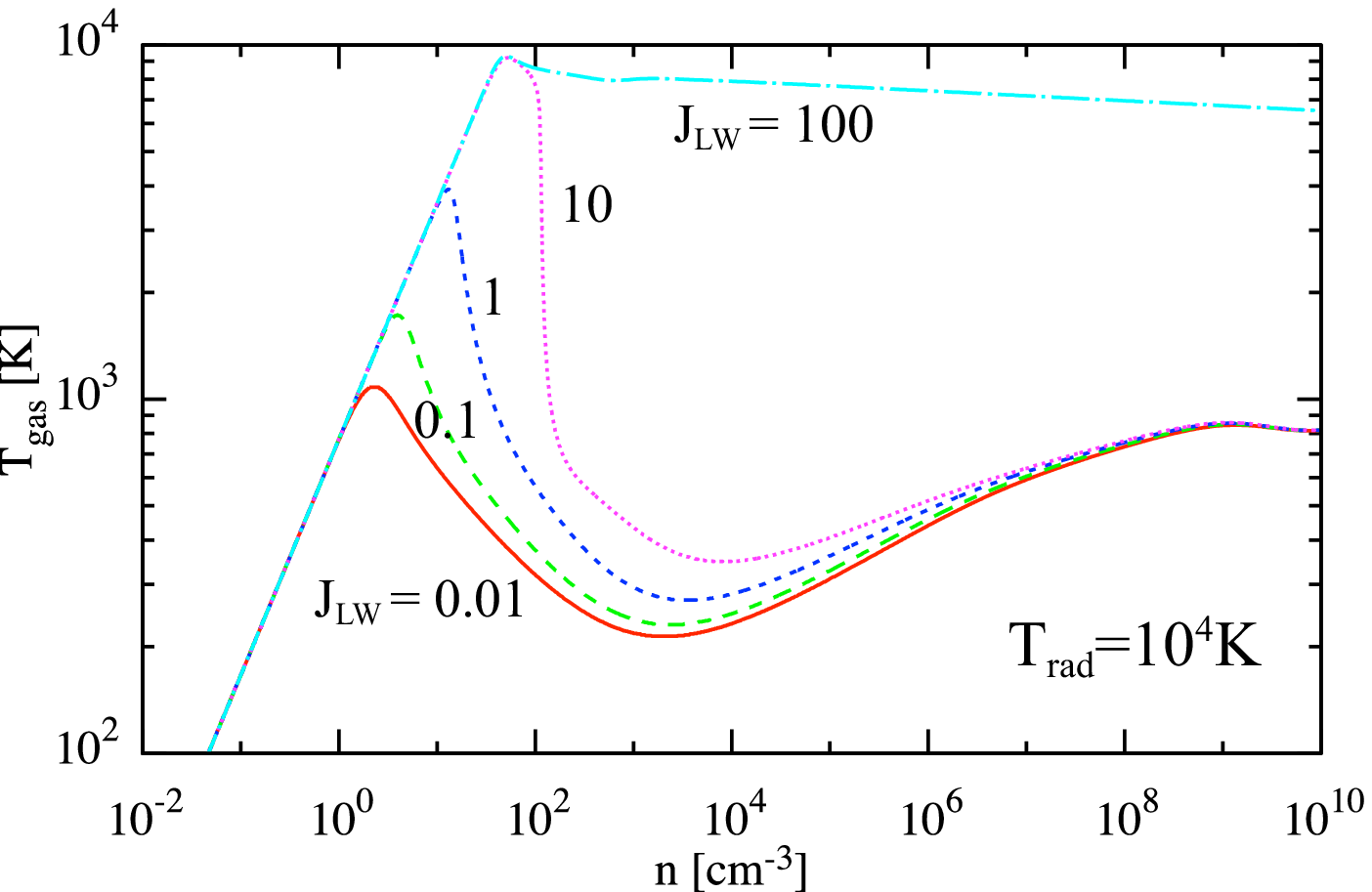} \hspace*{-0.3cm}
\includegraphics[width=9cm]{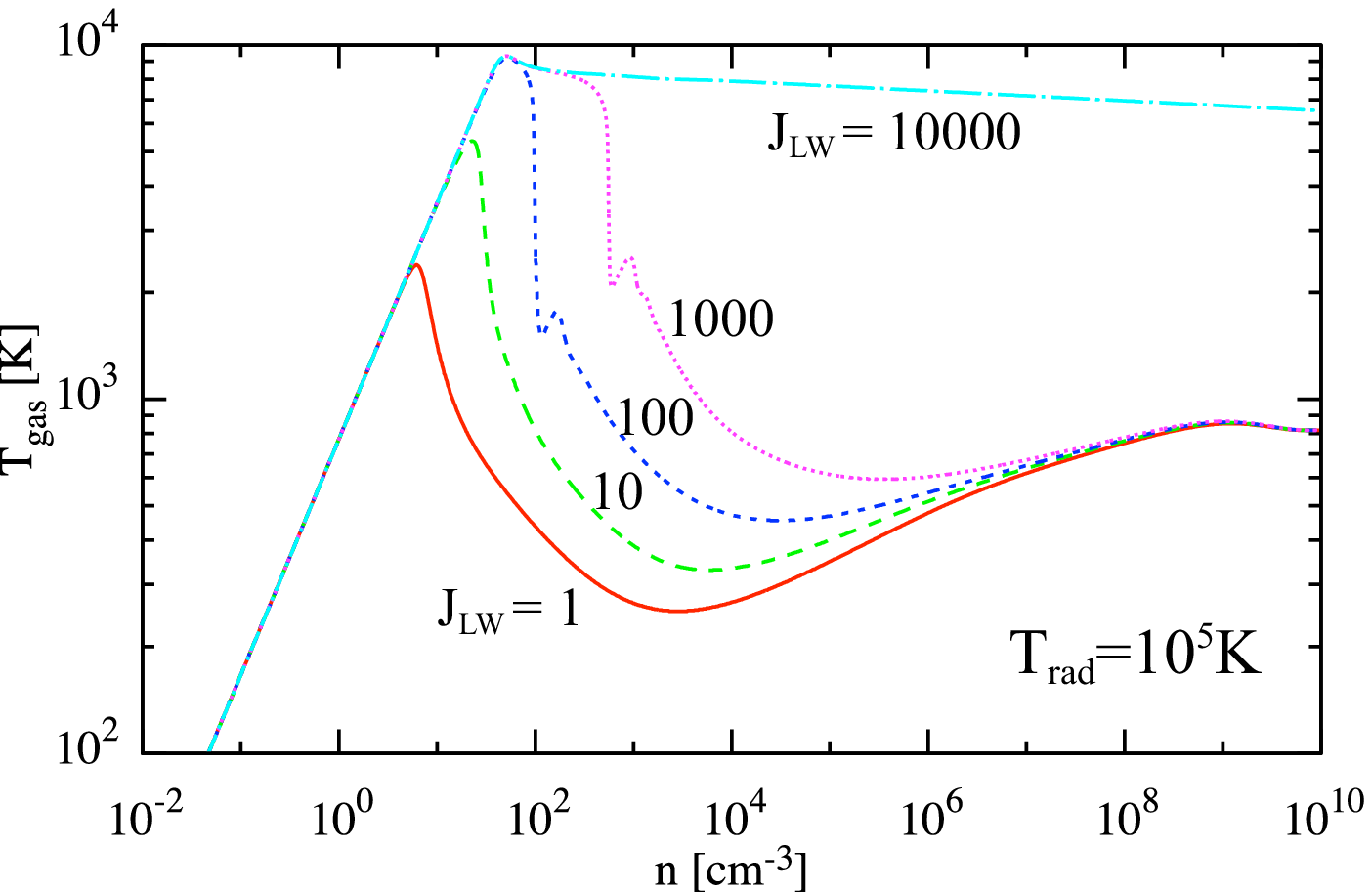}
\caption{Density-temperature relation for the collapse of primordial-gas
clouds under radiation with the black-body spectra with $T_{\rm rad}=10^4\,{\rm K}$
(top panel) and $10^5\,{\rm K}$ (bottom panel).  We take the
specific intensity at the LW bands as $J_{\rm
LW}=0.01,\,0.1,\,1,\,10,\,100$ and $J_{\rm
LW}=1,\,10,\,100,\,1000,\,10000$ for the $T_{\rm rad}=10^4\,{\rm K}$ and $10^5\,{\rm K}$ cases, respectively. } \label{fig:t1_4}
\end{figure}

The results are shown in Figs.~\ref{fig:t1_4}, where it can be clearly
seen that the evolutionary trajectories bifurcate to one of two types of
tracks: the atomic  and $\rm H_2$ cooling tracks.  In the case that the
atomic cooling track is chosen, the clouds evolve almost isothermally with $T_{\rm
gas}\sim 8000\,{\rm K}$ by atomic cooling.  On the other hand, in the
case that the $\rm H_2$ cooling track is chosen, the evolutionary trajectories
rapidly merge to the $\rm H_2$ cooling track when $\rm H_2$ cooling
becomes effective and the clouds cool down to $T_{\rm gas}\lesssim
1000\,{\rm K}$.  By increasing external radiation, the trajectories
get closer to the atomic cooling track, and finally merge to the
atomic cooling track at $J_{\rm LW,\,21}=100$ and $10000$ for $T_{\rm rad}=10^4\,{\rm K}$ and $10^5\,{\rm K}$,
respectively. The atomic cooling track is chosen
in the case that $\rm H_2$ molecules needed for $\rm H_2$ cooling
are suppressed by the strong external radiation.

The conditions of the gas when $n$ is close to the critical density
$n_{\rm cr}$ is crucial in determining which track is finally chosen
(O01).  Here, $n_{\rm cr}$ is defined as the  the density above which the population of the
internal states of $\rm H_2$ is determined by the LTE distribution due to sufficient
collisional excitation.
For both vibrational and rotational excitations of $\rm H_2$
molecules, $n_{\rm cr}$ is about $10^3 {\rm cm}^{-3}$ for $T_{\rm
gas}\sim 8000\,{\rm K}$.  The former and latter excitations are closely
related to the collisional dissociation and cooling processes,
respectively.  When $n>n_{\rm cr}$, $\rm H_2$ molecules are easily
dissociated via collisional dissociation and, in addition, cooling rate
per $\rm H_2$ molecule saturates and $\rm H_2$ cooling becomes less
effective than compressional heating.  On the other hand, the $\rm H_2$
formation channel given by Eqs. \eqref{eq:7} and \eqref{eq:8} becomes
effective as $n$ increases.  Thus, although $\rm H_2$ becomes easier to
be formed as $n$ increases until $n_{\rm cr}$, once a trajectory passes
through $n_{\rm cr}$, it is difficult to make transition from the atomic
cooling track to the $\rm H_2$ cooling track.  Therefore, the fate of a
trajectory can be known by examining whether the sufficient amount of
$\rm H_2$ molecules is formed around $n\sim n_{\rm cr}$.  Once $\rm H_2$
molecules are formed and $\rm H_2$ cooling becomes effective,
collisional dissociation is suppressed due to the decrease of $T_{\rm
gas}$ and, in addition, $\rm H_2$ photodissociation is suppressed due to
the self-shielding of the LW photons by $\rm H_2$ molecules, and thus
the trajectory rapidly converges to the $\rm H_2$ cooling track.

Let us see how the abundance of $\rm H_2$ is determined when the
formation and dissociation processes balance each other under external
radiation.  If the specific intensity of external radiation is about
that needed for the atomic cooling track, ${\rm H_2}$ photodissociation
is the main dissociation process around $n\sim n_{\rm cr}$ (O01; S10).
Thus, by equating the ${\rm H_2}$ formation rate of the channel
given by Eqs.~\eqref{eq:7} and \eqref{eq:8} and the $\rm H_2$ dissociation rate
of the photodissociation process given by Eq.~\eqref{eq:14}, we
obtain $k_{\rm form}^{\rm (eff)}\,n({\rm H})\,n({\rm e})=k_{\rm H_2,\,pd}\,n({\rm H_2})$.  By using this equation and Eqs.~\eqref{eq:10},
\eqref{eq:2} and \eqref{eq:22} and assuming $n({\rm H})\approx n$, we
obtain
\begin{align}
 y({\rm H_2})=
 \left[\frac{ k_{\rm form}^{(1)}}
 {\kappa_{\rm H_2,\,pd}\,J_{\rm LW}}\right]\,
 \left[\frac{k_{\rm form}^{(2)}\, n }{k_{\rm form}^{(2)}\, n+\kappa_{\rm H^-,\,pd}\,J_{\rm 2eV}}\right]\,
 n\, y(e)\,.
 \label{eq:9}
\end{align}
From this equation, it is clear that $y({\rm H_2})$ becomes small if
the first square bracket is suppressed due to large $\kappa_{\rm H_2,\,pd}\,J_{\rm LW}$ and/or the second is suppressed due to large
$\kappa_{\rm H^-,\,pd}\,J_{\rm 2eV}$.  In other words, the amount of
$\rm H_2$ can be suppressed by strong ${\rm H_2}$ and/or ${\rm H^-}$
photodissociation.

Equation~\eqref{eq:9} helps us to physically understand the $T_{\rm
rad}$ and $J_{\rm LW}$ dependence of the evolution.  To begin with, we
explain why the clouds under the radiation with the same $J_{\rm
LW}=100$ evolve along the atomic cooling track in the case  $T_{\rm
rad}=10^4\,{\rm K}$ but along the $\rm H_2$ cooling track in the case
$T_{\rm rad}=10^5\,{\rm K}$.  In the $T_{\rm rad}=10^4\,{\rm K}$ case, we
obtain $k_{\rm H^-,\,pd}\sim 5\times 10^{-6}{\rm s^{-1}}$ with
Eq.~\eqref{eq:3} and $k_{\rm form}^{(2)}\, n_{\rm cr}\sim 1\times
10^{-6} {\rm s^{-1}}$ with $n_{\rm cr}\sim 10^3 {\rm cm}^{-3}$ and
$k_{\rm form}^{(2)}=1.3\times10^{-9} {\rm cm^3 s^{-1}}$
(\cite{Glover:2008aa}).  Thus, the second square bracket of
Eq.~\eqref{eq:9} is significantly smaller than 1, meaning $\rm H^-$
photodissociation plays a role in suppressing $\rm H_2$ formation.  On the other
hand, in the $T_{\rm rad}=10^5\,{\rm K}$ case, $k_{\rm H^-,\,pd}\sim
1\times 10^{-9}$ and the second square bracket of Eq.~\eqref{eq:9} is
almost 1, meaning the effect of $\rm H^-$ photodissociation is
negligible.  Therefore, the two clouds with different $T_{\rm rad}$
evolve along different tracks although the strength of $\rm H_2$
photodissociation is same due to the same $J_{\rm LW}$.

Next, we explain why the clouds under the radiation with same $T_{\rm
rad}=10^5\,{\rm K}$ evolve along the atomic cooling track in the case
$J_{\rm LW}\leq1000$ but along the $\rm H_2$ cooling track in the
case $J_{\rm LW}=10000$.  Even in the case $J_{\rm LW}=10000$, the
second square-bracket of Eq.~\eqref{eq:9} is almost 1 and the effect of
$\rm H^-$ photodissociation is negligible.   In the case $J_{\rm LW}=10000$, however,
$\rm H_2$ photodissociation is very strong
and the first square-bracket of Eq.~\eqref{eq:9} becomes very small.
Therefore, the cloud under the radiation with $T_{\rm
rad}=10^5\,{\rm K}$ and $J_{\rm LW}=10000$ evolves along the atomic cooling track
by suppressing the $\rm H_2$ abundance with strong $\rm H_2$ photodissociation
without any help of $\rm H^-$ photodissociation.

\subsection{$J^{\rm crit}$ for black-body spectra}
\label{sec:t-dependence}

In this section, we present $J^{\rm crit}$ for the black-body spectra
with temperatures $7000\,{\rm K}<T_{\rm rad}<200000\,{\rm K}$, to
understand the dependence of $J^{\rm crit}$ on the hardness of the
spectrum of external radiation.  We calculate $J^{\rm crit}$ with
different forms of self-shielding factors because there is some
disagreement on the form of self-shielding factor, as mentioned in
Sec.~\ref{sec:UV_rad}.  We also make comparison of our results with the
literature (S10 and WG11).  In practice, $J^{\rm crit}$ is calculated
with the bisection method by examining whether $T_{\rm gas}$ is larger
or smaller than $4000\,{\rm K}$ at $n=10^7{\rm cm^{-3}}$.  In this
section, we use both $J_{\rm LW}$ and $J_{\rm Lyc}$ to specify the
strength of radiation, in order to make it easier to compare our result
with the literature ($J_{\rm Lyc}$ was used in, e.g., O01, S10, WG11).
Note that we mainly use $J_{\rm LW}$ other than this section because the
specific intensity at the LW bands is more directly related to the
physics we are interested in.

\begin{figure}
 \centering \hspace*{-0.3cm}
\includegraphics[width=9cm]{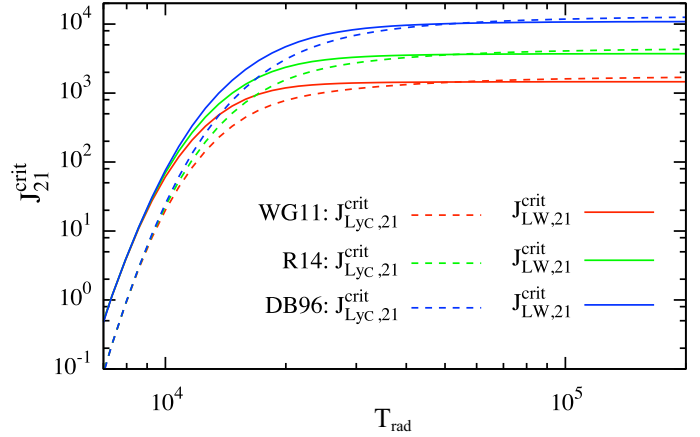} \caption{The
critical radiation intensity $J^{\rm crit}$ for the black-body spectra with
$T_{\rm rad}$. The critical LW intensity $J_{\rm LW,\,21}^{\rm crit}$ and 
the critical Lyc intensity $J_{\rm Lyc,\,21}^{\rm
crit}$ are plotted as solid and dashed lines, respectively.  The red,
green and blue lines are the results with the self-shielding factors of
WG11, R14 and DB96, respectively.  } \label{fig:t_jcrit}
\end{figure}

The results for our fiducial model, in which the self-shielding factor of
WG11 is used,
are shown as the red lines in Fig.~\ref{fig:t_jcrit}.
As $T_{\rm rad}$ increases, $J^{\rm crit}$ becomes larger:
$J_{\rm 21,\,LW}^{\rm crit}= 60$ at $T_{\rm rad}=10^4\,{\rm K}$,
$J_{\rm 21,\,LW}^{\rm crit}=1200$ at $T_{\rm rad}=2\times 10^4\,{\rm K}$.  
However, the
$T_{\rm rad}$ dependence of $J_{\rm 21,\,LW}^{\rm crit}$ becomes very
weak for the case $T_{\rm rad}\gtrsim 3\times 10^4\,{\rm K}$,
for which $J_{\rm 21,\,LW}^{\rm crit}\sim1400$ and is almost constant.

The $T_{\rm rad}$ dependence of $J^{\rm crit}$ can be understood with
Eq.~\eqref{eq:9} in a similar manner to that in
Sec.~\ref{sec:evolution}.  As $T_{\rm
rad}$ decreases $J^{\rm crit}$ becomes smaller because $\rm H^-$ photodissociation 
suppress $\rm H_2$ formation 
more effectively (the second square-bracket of Eq.~\eqref{eq:9} becomes
smaller).  For high $T_{\rm rad}$ ($T_{\rm rad}\gtrsim 3\times 10^4\,{\rm K}$),
the dependence of
$J_{\rm 21,\,LW}^{\rm crit}$ on $T_{\rm rad}$ is weak, because the effect of
$\rm H^-$ photodissociation is negligibly small (the second square-bracket of
Eq.~\eqref{eq:9} is almost unity).

\subsubsection{Influence of using different self-shielding factors}
\label{sec:fsh-dependence}

In the followings, we discuss the influence of using different forms of
self-shielding factors $f_{\rm sh}$. The form of $f_{\rm sh}$ derived in
\cite{Draine:1996aa} (hereafter DB96) has been widely used in the literature
(e.g. O01; S10). However, WG11 modified it to reproduce the results of
their radiative transfer calculations with three-dimensional hydrodynamic
simulations for gas with $T_{\rm gas}\sim 8000\,{\rm K}$.  Recently, R14
proposed another form of $f_{\rm sh}$, arguing that the form derived in
WG11 underestimates the strength of self-shielding compared with their
radiative transfer calculations with {\sc Cloudy} (\cite{Ferland:1998aa}).  

In this section, we introduce the three different forms of $f_{\rm sh}$.
First, the form of $f_{\rm sh}$ derived in WG11 is given by
Eq.~\eqref{eq:12}. Second, the form derived in DB96 is given by
\begin{align}
 f_{\rm sh}(N_{\rm H_2})=
\min\left[1,\,\left(\frac{N_{\rm H_2}}{10^{14}{\rm cm^{-2}}}\right)^{-3/4}\right]\,.
\label{eq:15}
\end{align}
Third, the form derived in R14 is given by
\begin{align}
  f_{\rm sh}(N_{\rm H_2}, T_{\rm gas})&=
\frac{1-\omega_{\rm H_2}(T_{\rm gas})}{(1+x'/b_5)^{\alpha(T_{\rm gas})}}\exp\left[-5\times10^{-7}(1+x')\right]\nnmb
&+\frac{\omega_{\rm H_2}(T_{\rm gas})}{(1+x')^{0.5}} \exp\left[-8.5\times10^{-4}(1+x')^{0.5}\right]\,,\nnmb
\label{eq:17}
\end{align}
where
\begin{align}
x'\equiv \frac{N_{\rm H_2}}{N_{\rm crit}(T_{\rm gas}) }\,,
\end{align}
\begin{align}
 \frac{N_{\rm crit}(T_{\rm gas})}{10^{14}\,{\rm cm^{-2}}}=
\begin{cases}
 1.3\left[1+\left(\frac{T_{\rm gas}}{600\,{\rm K}}\right)^{0.8}\right] & T_{\rm gas} < 3000\,{\rm K}\\
\left(\frac{T_{\rm gas}}{4760\,{\rm K}}\right)^{-3.8} & 3000\,{\rm K}\leq T_{\rm gas}< 4000\,{\rm K}\\
2.0&4000\,{\rm K}\leq T_{\rm gas}\qquad,
\end{cases}
\end{align}
\begin{align}
\omega_{\rm H_2}(T_{\rm gas})
= 0.013\left[1\!+\!\left(\frac{T_{\rm gas}}{2700\,{\rm K}}\right)^{\!1.3}\right]^{\frac{1}{1.3}}
\!\!\exp\left[-\!\left(\frac{T_{\rm gas}}{3900\,{\rm K}}\right)^{\!14.6}\right]\,,
\end{align}
 and 
\begin{align}
\alpha(T_{\rm gas}) 
=
\begin{cases}
 1.4 & T_{\rm gas} < 3000\,{\rm K}\\
\left(\frac{T_{\rm gas}}{4500\,{\rm K}}\right)^{-0.8} & 3000\,{\rm K}\leq T_{\rm gas}< 4000\,{\rm K}\\
1.1&4000\,{\rm K}\leq T_{\rm gas}\qquad.
\end{cases}
\end{align}

We present in Fig.~\ref{fig:t_jcrit} the critical intensity $J^{\rm crit}$ for the black-body
radiation with $T_{\rm rad}$ for the three forms of $f_{\rm
sh}$ given above. Difference in  $f_{\rm
sh}$ affects $J^{\rm crit}$ more for higher $T_{\rm rad}$.  For instance, for the $T_{\rm
rad}=10^4\,{\rm K}$ cases, in which $\rm H^-$ photodissociation plays a
key role in suppressing the $\rm H_2$ formation independently of self-shielding, 
$J^{\rm crit}$ is almost same for all three cases.
 On the other
hand, for the $T_{\rm rad}=10^5\,{\rm K}$ cases, in which the evolutionary
trajectories are totally determined by the strength of $\rm H_2$
photodissociation, difference in $f_{\rm sh}$ affects significantly
and $J_{\rm 21,\,LW}^{\rm crit}=1500\,,3700$ and
11000 with $f_{\rm sh}$ of WG11, R14 and DB96, respectively.  In
summary, the influence of using different forms of $f_{\rm sh}$ is not
negligible.

Before ending this section, we make comparison of our results with
the literature (S10 and WG11).  S10 used $f_{\rm sh}$ of DB96 to
obtain $J_{\rm 21,\,Lyc}^{\rm crit}=39$ and $12000$ for $T_{\rm
rad}=10^4\,{\rm K}$ and $10^5\,{\rm K}$, respectively.  Our results
with the same $f_{\rm sh}$ are $J_{\rm 21,\,Lyc}^{\rm
crit}=25$ and $14000$ for $T_{\rm rad}=10^4\,{\rm K}$ and $10^5\,{\rm K}$, respectively.  WG11 used their own $f_{\rm sh}$ to
obtain $J_{\rm 21,\,Lyc}^{\rm crit}=1400$ for the black-body spectrum with
$T_{\rm rad}=10^5\,{\rm K}$.  Our result with the same $f_{\rm
sh}$ is $J_{\rm 21,\,Lyc}^{\rm crit}=1600$ for $T_{\rm
rad}=10^4\,{\rm K}$.  In general, our results are in good agreement with those
obtained in S10 and WG11. The remaining differences might be due to
the use of different chemical reaction rates and cooling functions,
because we use those of \cite{Glover:2008aa} while S10 and WG11 used those
of \cite{Galli:1998aa}.

\subsection{$J^{\rm crit}$ for realistic spectra}
\label{sec:realistic}
In this section, we present $J^{\rm crit}$ for realistic spectra
considering various models of source galaxies and IGM radiative
transfer.  It is needed to obtain $J^{\rm crit}$ for each realistic
spectrum, because realistic spectra in general do not look like the
black-body spectra and cannot be parameterized with a single parameter
like $T_{\rm rad}$.  To obtain the spectra of galaxies, we adopt the
spectral model by \cite{Inoue:2011aa}, who added nebular lines and continua to
the stellar population synthesis models of \cite{Schaerer:2003aa}
for $Z/Z_\odot=0$ (Pop III)
and $Z/Z_\odot=5\times 10^{-4}$
and of {\sc Starburst99} (\cite{Leitherer:1999aa}) for
$Z/Z_\odot=0.02$ and $Z/Z_\odot=0.2$ by a metallicity dependent way.

\begin{table}
 \caption{Galaxy/IGM models explored}
\label{tab:model}
 \begin{tabular}{ll} \hline\\[-5pt]
IMF& Salpeter IMF with 1-100$M_\odot$ (fixed)\!\!\!\!\!\!\!\!\\
metallicity $(Z/Z_\odot)$&
$0$ (Pop III), $5\!\times\! 10^{-4}$, $0.02$, $0.2$\\
\smash[b]{\lower14pt\hbox{SF type: Age}}&
 instantaneous starburst (IS):\\
& \hspace*{0.5cm}1\,\,Myr, 10\,Myr, 100\,Myr-1\,Gyr$^*$\\
  & constant star formation (CS):\\
&\hspace*{0.5cm} 1\,Myr, 10\,Myr, 100\,Myr, 500\,Myr\!\!\!\!\!\! \\
escape fraction ($f_{\rm esc}$)& 0,\, 0.5\\[3pt]
\end{tabular}
 \begin{tabular}{ll}
\hline \\[-5pt]
IGM absorption (Ly$\alpha$) & complete/no absorption\hspace*{1.2cm}\\ 
IGM absorption ($>$Lyc)& complete absorption (fixed)\\[3pt]\hline
\end{tabular}\\
$^* $ IS galaxies with age between 100\,Myr and 1\,Gyr 
are studied with the bin width of 0.1 on a logarithmic scale.
\end{table}

\begin{figure}
 \centering
 \hspace*{-0.3cm}\includegraphics[width=9cm]{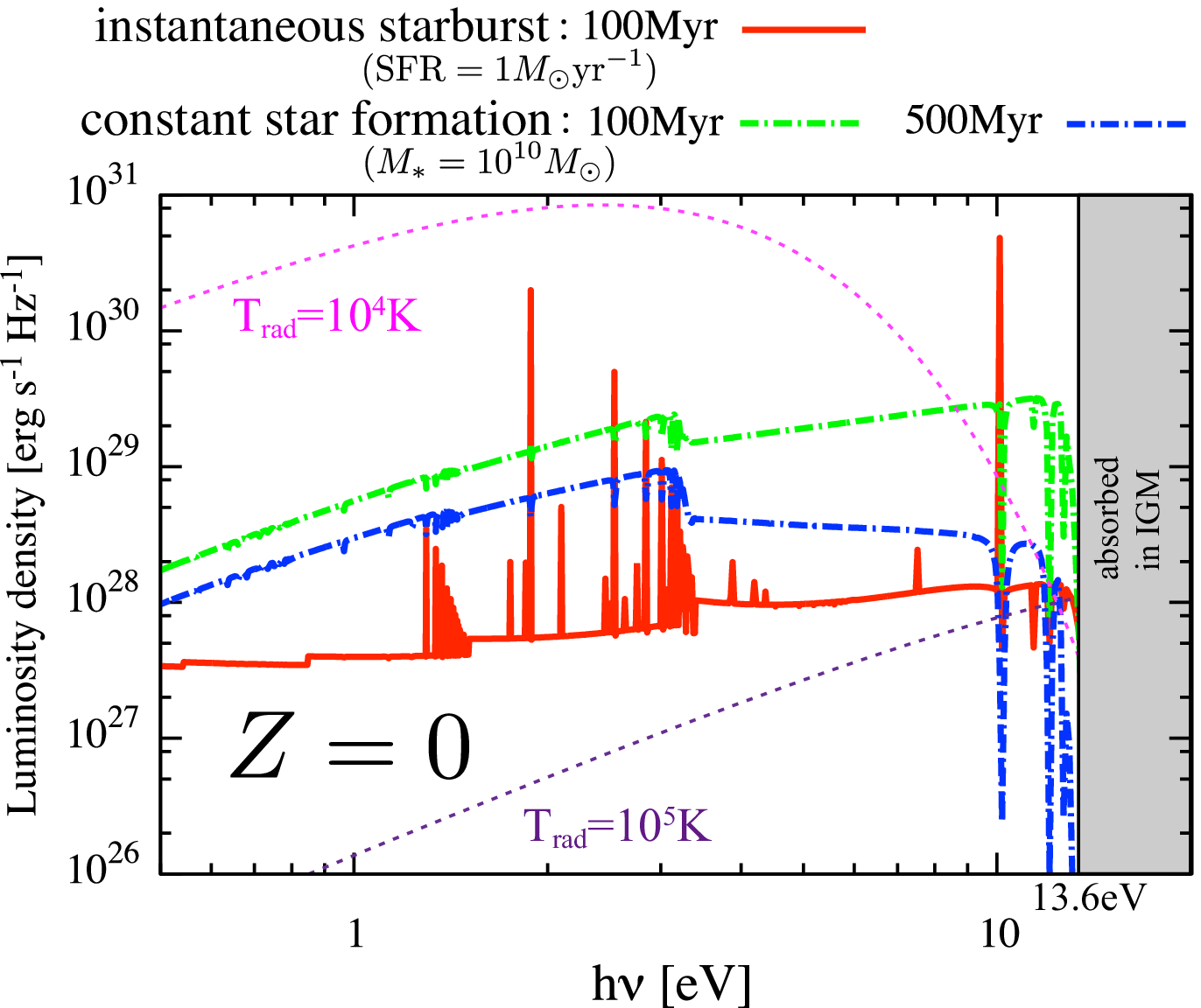}\\
\vspace*{1cm}
 \hspace*{-0.3cm}\includegraphics[width=9cm]{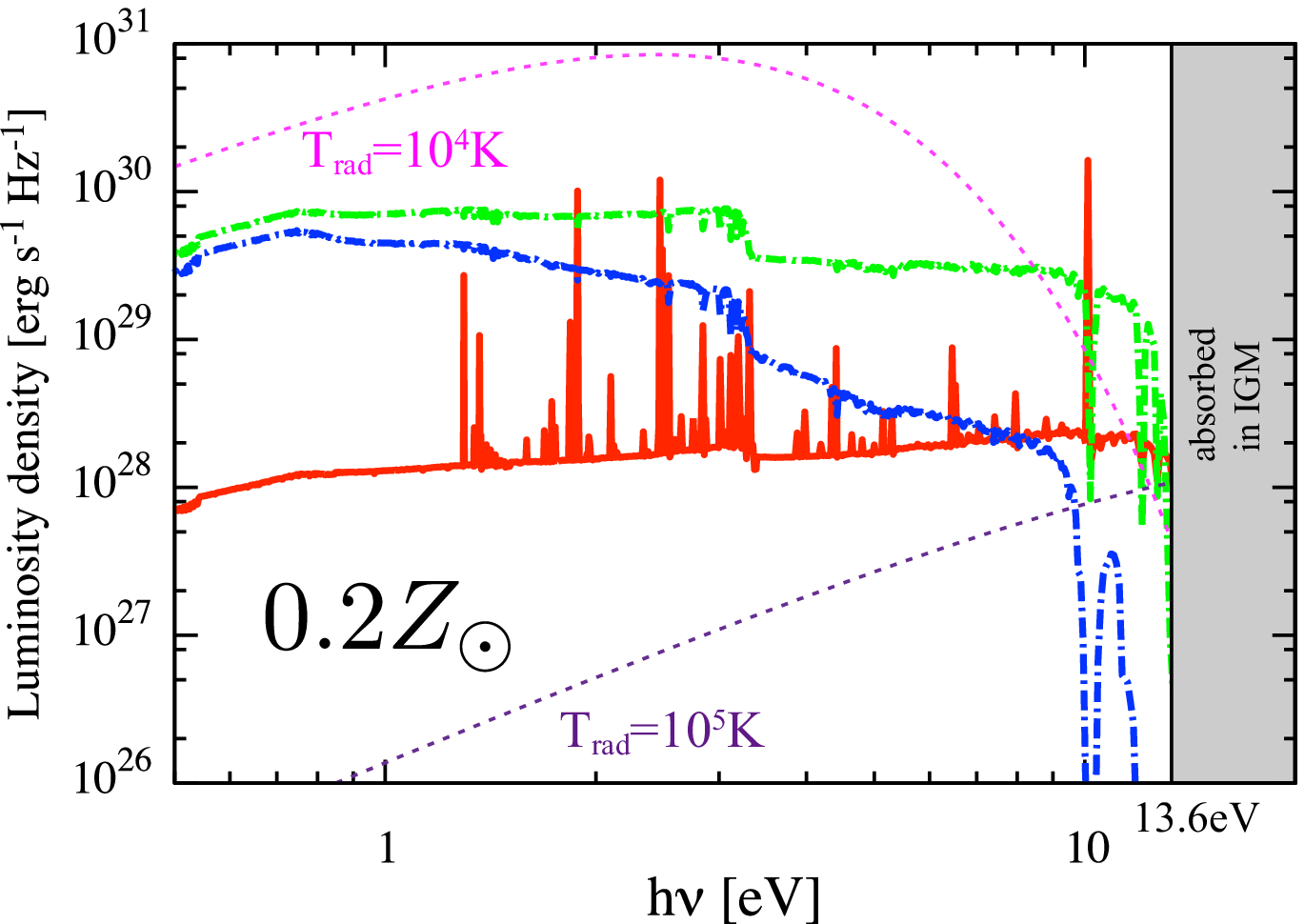}
 \caption{The spectra of galaxies with metallicity $Z=0$ (Pop III) (top
 panel) and $Z/Z_\odot=0.2$ (bottom panel).  For both cases, we take
 $f_{\rm esc}=0$ and plot the spectra of the constant star formation
 galaxies with a duration of star formation 100\,Myr (with the star
 formation rate (SFR)= $1M_\odot {\rm yr^{-1}}$) and the instantaneous
 starburst galaxies with a time since the burst 100\,Myr and 500\,Myr (with
 the total stellar mass $M_{*}=10^{10}M_\odot $).  We do not consider
 Ly$\alpha$ absorption by the IGM in this figure.  The emission-line
 width is assumed to be 300 km s$^{-1}$ for illustrating purposes.  For
 comparison, we also plot the black-body spectra with  $T_{\rm rad}=10^4\,{\rm K}$ and  $10^5\,{\rm K}$
 arbitrary scaled in the panels.
} \label{fig:pop3_spec}
\end{figure}

The models of galaxies and IGM radiative transfer explored in this paper
are summarized in Table \ref{tab:model}. Since we are interested in the
early universe, we focus on metal-poor galaxies with $Z/Z_\odot=0$ (Pop
III), $5\times 10^{-4}$, $0.02$ and $0.2$,
where the solar metallicity $Z_\odot=0.02$.   We
assume a Salpeter type initial mass function (IMF) with the stellar mass range $1 - 100M_\odot$.  We consider
two types of star formation (SF) histories: instantaneous starburst (IS) and
constant star formation (CS).  For the CS galaxies
we assume the duration of star formation
is 1\,Myr, 10\,Myr, 100\,Myr and 500\,Myr, while for the IS galaxies
we assume the time since the burst is 1\,Myr, 10\,Myr and  between
100\,Myr and 1\,Gyr with the bin width of 0.1 on a logarithmic
scale.  We consider the two cases for the escape fraction of ionizing photons from source
galaxies $f_{\rm esc}$ and take $f_{\rm esc}=$ 0 and 0.5, where the absorbed energies of the
ionizing photons are converted to the nebular emission.  We assume that all
ionizing photons from galaxies are absorbed by the intergalactic
medium (IGM), but consider the two cases for the Ly$\alpha$ line, where
it is either completely absorbed by the IGM or not at all.
As a whole, we explore $4\times4\times2\times2=64$ and
$4\times13\times2\times2=208$ models of the IS and CS galaxies,
respectively. 
It should be noted that old galaxies studied in this
section are not proper candidates for sources of radiation contributing to DCBH
formation at $z\gtrsim10$, since the age of the universe is about 500Myr
at $z=10$.  However, we explore a wide
range of galaxies to see the dependence of
$J^{\rm crit}$ on spectra clearly.

As examples, we show the spectra of
the $Z=0$ (Pop III) and $Z/Z_\odot=0.2$ galaxies in Figs.~\ref{fig:pop3_spec}.  The
spectra are roughly flat in the frequency range $h\nu\lesssim 10{\rm
eV}$ due to the superposition of the stellar emission with various
effective temperatures. The number of the LW photons from the old IS
galaxies is exponentially suppressed, because the high temperature stars
that contribute to producing the LW photons no longer emit radiation in
such galaxies due to their short lifetimes.

\begin{figure}
 \centering
 \hspace*{-0.3cm}\includegraphics[width=9cm]{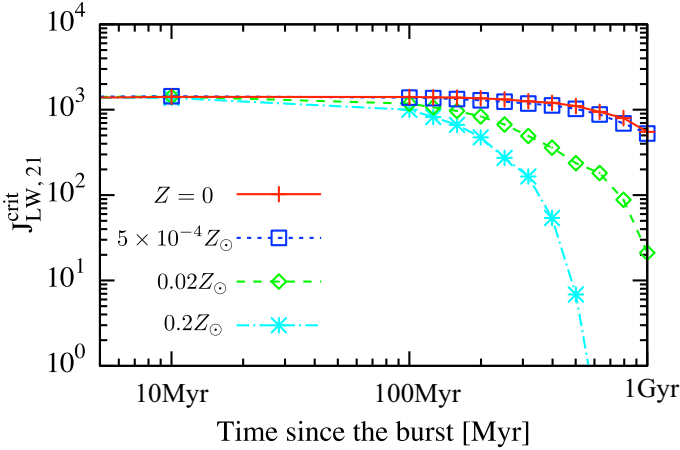}
 \caption{The critical LW intensity $J_{\rm LW,\,21}^{\rm crit}$ for
 realistic spectra of the IS galaxies with $Z=0$,
 $5\times 10^{-4}Z_\odot$, $0.02Z_\odot$ and $0.2Z_\odot$.  
We  assume complete Ly$\alpha$ absorption and $f_{\rm esc}=0$.  The
 horizontal axis is the time since the burst.  $J_{\rm LW,\,21}^{\rm
 crit}\sim1400$ at 1\,Myr since the burst irrespective of metallicity. 
In the case of the CS galaxies,
$J_{\rm LW,\,21}^{\rm crit}=1300-1400$ irrespective of the metallicity and the duration of SF.
 }
 \label{fig:real_jcrit}
\end{figure}

The results of $J^{\rm crit}$ for realistic spectra are summarized as
follows.  While $J_{\rm LW,\,21}^{\rm crit}=1300-1400$ for all CS
galaxies irrespective of the metallicity and duration of SF, $J^{\rm
crit}$ has a wide range of values for the IS galaxies
depending on the models, as shown in Fig.~\ref{fig:real_jcrit}.  We plot
only the cases of complete Ly$\alpha$ absorption and $f_{\rm esc}=0$ in
Fig.~\ref{fig:real_jcrit}, because the effects of changing Ly$\alpha$
absorption and $f_{\rm esc}$ make at most 5\% difference to the values
of $J^{\rm crit}$ although the Ly$\alpha$ line and the nebular emission
contribute to $\rm H^-$ photodissociation and slightly reduce the value of $J^{\rm
crit}$.
The critical intensity $J^{\rm crit}$ decreases as
the IS galaxies become older or more metal-enriched, although the dependence
is weak for the young or extremely metal-poor
galaxies. For the young galaxies with
the time since burst less than 100\,Myr
$J_{\rm LW,\,21}^{\rm crit}= 1000-1400$,
while for the extremely metal-poor and not very old galaxies ($Z\leq 5\times 10^{-4}Z_\odot$
and the time since burst is less than 500\,Myr)
$J_{\rm LW,\,21}^{\rm crit}\approx 1400$ and is almost
constant.

\subsection{The reason for the dependence of $J^{\rm crit}$ on spectra}
\label{sec:fitting-formula}
In this section, we explain the reason for the dependence of $J^{\rm
crit}$ on the spectral shape of external radiation, as seen in Sec.~\ref{sec:t-dependence} and
Sec.~\ref{sec:realistic}, by pointing out the key parameter determining
$J^{\rm crit}$.  We then develop a method to
estimate $J^{\rm crit}$ for a given spectrum without calculating
the evolution of the clouds.
\subsubsection{The key parameter determining $J^{\rm crit}$}
\label{sec:kph-dependence}

\begin{table*}
\begin{minipage}{14cm}
\caption{The relation between $T_{\rm rad}$ and $k_{\rm H^-,\,pd}/k_{\rm H_2,\,pd}$.}  \label{tab:T_rkph}
 \begin{tabular}{lcccccccc} \hline \\[-5pt]
  $T_{\rm rad}\,$[K]
  &$8\times 10^3$&$1\times 10^4$&$2\times 10^4$&$3\times10^4$
  &$5\times 10^4$&$1\times 10^5$&$2\times 10^5$\\[4pt] \hline\\[-5pt]
  $k_{\rm H^-,\,pd}/k_{\rm H_2,\,pd}$
  &$8.7\times10^5$&$4.6\times10^4$&$2.1\times 10^2$&$4.6\times 10^1$
		  &$1.7\times 10^1$ &$1.0\times 10^1$ &$8.1\times 10^0$ \\[5pt] \hline
 \end{tabular}
\end{minipage}
\end{table*}

\begin{table*}
\begin{minipage}{9cm}
 \caption{The values of $\alpha_{\rm H^-,\,pd}$ for the power-law spectra
$J(\nu)\propto \nu^{s}$}  \label{tab:aHm_pow} \hspace*{-0.5cm}
 \begin{tabular}{crrrrrrrrr} \hline \\[-5pt]
  $s$&-2&-1.5&-1&-0.5&0&0.5&1&1.5&2\\[3pt]\hline\\[-5pt]
$\alpha_{\rm H^-,\,pd}$
 &1.44&1.21&1.06&0.99&1.00&1.12&1.41&2.03&3.32\\[5pt] \hline
 \end{tabular}
 \end{minipage}
\end{table*}

In this section, we propose a hypothesis that the ratio of the ${\rm
H^-}$ and ${\rm H_2}$ photodissociation rates, $k_{\rm H^-,\,pd}/k_{\rm H_2,\,pd}$, is the key parameter determining the dependence of $J^{\rm
crit}$ on spectra, and prove its validity in the followings.  We
come up with this hypothesis because, as explained in
Sec.~\ref{sec:evolution}, in the cases strong ${\rm H^-}$
photodissociation suppresses ${\rm H_2}$ formation, smaller $J_{\rm LW}$
(and hence weaker ${\rm H_2}$ photodissociation) is needed to suppress
${\rm H_2}$ cooling.  In this section, the quantity written as $k_{\rm H_2,\,pd}$ is not the true value realized in clouds during the evolution but
that defined by Eq.~\eqref{eq:2} without considering the effect of self-shielding.
Here, we are interested in the quantity directly related to external radiation.

\begin{figure}
 \hspace*{-0.3cm}
\includegraphics[width=9cm]{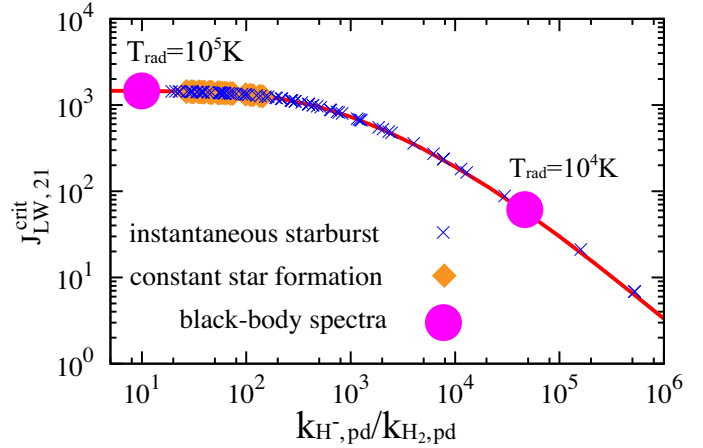} \caption{The
critical LW intensity $J_{\rm LW,\,21}^{\rm crit}$ with respect to
$k_{\rm H^-,\,pd}/k_{\rm H_2,\,pd}$ for the realistic spectra (blue
points and orange diamonds corresponding to IS and CS galaxies,
respectively) and black-body spectra with $T_{\rm rad}=10^4\,{\rm K}$
and $10^5\,{\rm K}$ (magenta dots).  
The value of $k_{\rm H^-,\,pd}/k_{\rm H_2,\,pd}$ represents the hardness of each spectrum.  We
 also plot $J_{\rm LW,\,21}^{\rm crit}$ with respect to $k_{\rm H^-,\,pd}/k_{\rm H_2,\,pd}$ obtained by assuming $k_{\rm H_2^+,\,pd}/k_{\rm H^-,\,pd}=0.1$ (red line).}  \label{fig:kphr_jcrit}
\end{figure}

To demonstrate that $k_{\rm H^-,\,pd}/k_{\rm H_2,\,pd}$ is the only
parameter determining $J^{\rm crit}$, we obtain $k_{\rm H^-,\,pd}/k_{\rm
H_2,\,pd}$ for the realistic and black-body spectra studied in this
paper and plot them with $J^{\rm crit}$ in Fig.~\ref{fig:kphr_jcrit}.
It is clear from Fig.~\ref{fig:kphr_jcrit} that there is one-to-one
correspondence between $k_{\rm H^-,\,pd}/k_{\rm H_2,\,pd}$ and $J^{\rm
crit}$.  In other words, $J^{\rm crit}$ is solely determined by $k_{\rm
H^-,\,pd}/k_{\rm H_2,\,pd}$.  We plot only the results for the complete
Ly$\alpha$ absorption and $f_{\rm esc}=0$ cases in
Fig.~\ref{fig:kphr_jcrit}, because the effects of changing Ly$\alpha$
absorption and $f_{\rm esc}$ make little difference.  The $k_{\rm
H^-,\,pd}/k_{\rm H_2,\,pd}$ dependence of $J^{\rm crit}$ can be
understood with Eq.~\eqref{eq:9}, in the same manner as the $T_{\rm
rad}$ dependence (see the last part of Sec.~\ref{sec:t-dependence}).

Note that $k_{\rm H^-,\,pd}/k_{\rm H_2,\,pd}$ can be
regarded as a proxy for the hardness of the spectrum.  The
relation between $T_{\rm rad}$ and $k_{\rm H^-,\,pd}/k_{\rm H_2,\,pd}$
are given in Table \ref{tab:T_rkph}. The ratio $k_{\rm H^-,\,pd}/k_{\rm H_2,\,pd}$
increases as $T_{\rm rad}$ decreases, corresponding to the fact that
${\rm H^-}$ photodissociation becomes more effective as the spectrum
becomes soft. The dependence of $k_{\rm H^-,\,pd}/k_{\rm H_2,\,pd}$ on
$T_{\rm rad}$ becomes weak for high $T_{\rm rad}$ ($\gtrsim
5\times 10^5\,{\rm K}$). This is because for such high $T_{\rm rad}$ the
spectrum obeys the Rayleigh-Jeans law ($J(\nu)\propto \nu^2$) in the frequency range
contributing to ${\rm H^-}$ and ${\rm H_2}$ photodissociation ($0.76\,{\rm eV}<h\nu < 13.6\,{\rm eV}$).

By obtaining the relation between $k_{\rm H^-,\,pd}/k_{\rm H_2,\,pd}$
and $J^{\rm crit}$ in advance, $J^{\rm crit}$ for a given spectrum can
be estimated from this relation without calculating the evolution of
the clouds.  However, there remains one uncertainty. 
In order to determine the evolution of clouds, the ${\rm
H_2^+}$ photodissociation rate should be specified in addition to
$J_{\rm LW}$ (which determines $k_{\rm H_2,\,pd}$ by Eq.~\eqref{eq:2})
and $k_{\rm H^-,\,pd}/k_{\rm H_2,\,pd}$.  In the followings, we assume
$k_{\rm H_2^+,\,pd}/k_{\rm H^-,\,pd}=0.1$, motivated by the fact that
$k_{\rm H_2^+,\,pd}/k_{\rm H^-,\,pd}\sim 0.1$ when $T_{\rm
gas}\sim8000{\rm K}$ during the evolution of the clouds under the external
radiation with the realistic and thermal spectra.  This assumption is
further justified by the fact that the detailed value of $k_{\rm H_2^+,\,pd}/k_{\rm H^-,\,pd}$ is not important.  We have also made calculations 
for the cases with $k_{\rm H_2^+,\,pd}$ 10 times larger or smaller than the
true values, but have found only negligible difference in the results.  The
relation between $J^{\rm crit}$ and $k_{\rm H^-,\,pd}/k_{\rm H_2,\,pd}$ obtained with the above assumption is shown as the red line in
Fig.~\ref{fig:kphr_jcrit}.  For the realistic and thermal spectra studied in
this paper, this relation almost perfectly reproduces the $J^{\rm crit}$
from the information of $k_{\rm H^-,\,pd}/k_{\rm H_2,\,pd}$.

For later convenience, we present a fitting formula to the relation given above,
 \begin{align}
J_{\rm LW,\,21}^{\rm crit} &= 
\begin{cases}
\ds
1400&
x\leq 0		 \\
\ds
1400
\times 
10^{\left(a_1 x+ a_2 x^2\right)}
&	
x> 0
	\end{cases}\,,
 \label{eq:4}
 \end{align}
where 
\begin{align}
 x&=\log_{10}\left(k_{\rm H^-,\,pd}/k_{\rm H_2,\,pd}\right)-2\,,
\label{eq:19}
\end{align}
and
\begin{align}
 a_1&=-0.19\,,\quad 
a_2=-0.12\,.\quad
\label{eq:23}
\end{align}
 We find, for the realistic and
black-body spectra studied in this paper, this fitting formula
reproduces the $J^{\rm crit}$ from $k_{\rm H^-,\,pd}/k_{\rm H_2,\,pd}$
with at most 10\% error in the range $1< k_{\rm H^-,\,pd}/k_{\rm H_2,\,pd}<
10^5$.  With this fitting formula, $J^{\rm crit}$ can be easily
estimated from $k_{\rm H^-,\,pd}/k_{\rm H_2,\,pd}$ for a given spectrum.

\subsubsection{A method to estimate $J^{\rm crit}$ for a given spectrum}
\label{sec:JH--dependence}

Here, we propose a simple and easy method to estimate $J^{\rm crit}$ for
a given spectrum.  As mentioned in Sec.~\ref{sec:kph-dependence}, the
relation between $k_{\rm H^-,\,pd}/k_{\rm H_2,\,pd}$ and $J^{\rm
crit}$ can be used to estimate $J^{\rm crit}$.
However, to use this relation, the
frequency integral in Eq.~\eqref{eq:3} is needed to be evaluated in
obtaining $k_{\rm H^-,\,pd}$.  The information of the spectral shape
is contained in $\kappa_{\rm H^-,\,pd}$ in Eq.~\eqref{eq:22},
which can be parameterized with $\alpha_{\rm H^-,\,pd}$ as
\begin{align}
\kappa_{\rm H^-,\,pd}=\alpha_{\rm H^-,\,pd}\kappa_{\rm H^-,\,pd}^{(0)}\,,
\label{eq:1} 
\end{align}
where $\kappa_{\rm H^-,\,pd}^{(0)}=1.1\times 10^{11}$ (in cgs unit) is
defined as $\kappa_{\rm H^-,\,pd}$ for the flat spectrum ($J(\nu)={\rm
const.}$) The values of $\alpha_{\rm H^-,\,pd}$ for the power-law
spectra $J(\nu)\propto \nu^{s}$ are given in Table~\ref{tab:aHm_pow}.
In the followings, we avoid the numerical integration of
Eq.~\eqref{eq:3} by approximating $\kappa_{\rm H^-,\,pd}$ with
$\kappa_{\rm H^-,\,pd}^{(0)}$. The realistic spectra are roughly flat in
the frequency range $h\nu\lesssim 10\,{\rm eV}$, as mentioned in
Sec.~\ref{sec:realistic}, and thus the error due to this approximation
can be estimated with the spread of $\alpha_{\rm H^-,\,pd}$ around a
flat spectrum and is expected to be small.
By using this approximation with Eqs.~\eqref{eq:2} and \eqref{eq:22}, we
can estimate $k_{\rm H^-,\,pd}/k_{\rm H_2,\,pd}$ from $J_{\rm
2eV}/J_{\rm LW}$ simply as
\begin{align}
 k_{\rm H^-,\,pd}/k_{\rm H_2,\,pd}\approx 79\, J_{\rm 2eV}/J_{\rm LW}.
 \label{eq:5}
\end{align}

In the light of this relation, we modify the fitting formula
given by Eqs.~\eqref{eq:4}, \eqref{eq:19} and \eqref{eq:23} 
by redefining $x$ in Eq.~\eqref{eq:19} as
 \begin{align}
x&=\log_{10}\left(79\, J_{\rm 2eV}/J_{\rm LW}\right)-2\,.
 \label{eq:6}
 \end{align}
To check the validity of the formula given by
Eqs.~\eqref{eq:4}, \eqref{eq:23} and \eqref{eq:6}, we obtain $J_{\rm
2eV}/J_{\rm LW}$ for the realistic and black-body spectra studied in this paper
and compare them with $J^{\rm crit}$ estimated from $J_{\rm 2eV}/J_{\rm LW}$ with this formula.  The result is
that the formula reproduces $J^{\rm crit}$
with at most 30\% error for both the realistic and black-body spectra.  Although the
error is larger than the formula given in Sec.~\ref{sec:kph-dependence},
it is still practically negligible considering an order-of-magnitude
scatter of $J^{\rm crit}$ due to the diversity in the three-dimensional structure of
the clouds (S10; \cite{Latif:2014ab}).

\section{Summary and discussion}
\label{sec:summary-discussion}
By using the one-zone model described in Sec.~\ref{sec:model}, we have
calculated the critical intensity of external radiation needed for primordial-gas clouds in 
halos with $T_{\rm
vir}\gtrsim 10^4\,{\rm K}$
to form DCBHs by suppressing $\rm H_2$ cooling.  By performing series of calculations for 
various types of external radiation, we have examined the dependence
of $J^{\rm crit}$ on the spectral shape of external radiation.

In Sec.~\ref{sec:t-dependence}, we have seen how $J^{\rm crit}$ changes
depending on the temperature of the black-body spectra between $7000\,{\rm
K}< T_{\rm rad}<20000\,{\rm K}$.  In Sec.~\ref{sec:realistic}, we have
determined $J^{\rm crit}$ for the realistic spectra of the metal-poor galaxies,
by taking the data from the stellar population synthesis models. 
We have found $J^{\rm crit}$ is not sensitive to the age
or metallicity for the constant star formation galaxies with $J_{\rm
LW,\,21}^{\rm crit}=1300-1400$, while $J^{\rm crit}$ decreases as
galaxies become older or more metal-enriched for the instantaneous
starburst galaxies.  However, such dependence for the instantaneous
starburst galaxies is weak for the young or extremely metal-poor
galaxies: $J_{\rm LW,\,21}^{\rm crit}= 1000-1400$ for the young (the age
less than 100 Myr) galaxies and $J_{\rm LW,\,21}^{\rm crit}\approx 1400$
for the extremely metal-poor ($Z<5\times 10^{-4}Z_\odot$) 
and not very old (the age less than 500 Myr) galaxies. 
It should be noted that the above values of $J^{\rm
crit}$ are obtained with $f_{\rm sh}$ of WG11 and that those obtained
with $f_{\rm sh}$ of R14 are about two times larger than the above
values, as shown in Sec.~\ref{sec:fsh-dependence}.  It is important to
precisely determine the form of $f_{\rm sh}$ but is beyond the scope of
this work.

We have also found that the dependence of $J^{\rm crit}$ on the spectral shape is
totally attributable to a single parameter $k_{\rm H^-,\,pd}/k_{\rm H_2,\,pd}$ in
Sec.~\ref{sec:kph-dependence}.  By using the one-to-one correspondence
between $k_{\rm H^-,\,pd}/k_{\rm H_2,\,pd}$ and $J^{\rm crit}$ and the
approximate relation between $k_{\rm H^-,\,pd}$ and $J_{\rm 2eV}$, we
have proposed a formula given by Eqs.~\eqref{eq:4}, \eqref{eq:23} and
\eqref{eq:6} to estimate $J^{\rm crit}$.  With this formula, $J_{\rm
crit}$ is reproduced with at most 30\% error
from the information of $J_{\rm 2eV}/J_{\rm LW}$
for the realistic and black-body spectra studied in this
paper.

Let us discuss the implication of our results.  In the followings, we
adopt $J_{\rm LW,\,21}^{\rm crit}=1400$ as the fiducial value, 
because it is the typical value for young and metal-poor
galaxies commonly present in the supposed DCBH
formation era at $z\gtrsim 10$. It has been known that the value of $J^{\rm
crit}$ is much higher than the averaged cosmic LW background $J_{\rm
bg,\,LW,\,21}\lesssim 0.1$ in the whole history of the Universe (see,
e.g., \cite{OShea:2008aa,Johnson:2013ab}), and thus the clouds
need to be irradiated by unusually nearby and/or strong sources 
to achieve $J_{\rm LW}> J_{\rm LW}^{\rm crit}$.  By performing
semi-analytical computations with N-body simulations (A12) and
Monte-Carlo simulations (D14), A12 and D14 estimated the DCBH number
density $n_{\rm DCBH}$ with assumption that DCBHs are formed in all the atomic-cooling
halos with $J_{\rm LW}> J_{\rm LW}^{\rm crit}$.  
However, they may have overestimated $n_{\rm DCBH}$ due to
the smaller values of $J^{\rm crit}$ used in their estimation ($J_{\rm
LW,\,21}^{\rm crit}=30$ and 300 are used in A12 and D14, respectively).

We re-estimate $n_{\rm DCBH}$ with our $J_{\rm LW,\,21}^{\rm crit}=1400$
by extrapolating the results of A12 and D14. The estimate of $n_{\rm
DCBH}$ changes from $n_{\rm DCBH}\sim 10^{-7}{\rm cMpc^{-3}}$ to $n_{\rm
DCBH}\sim 10^{-10}{\rm cMpc^{-3}}$ at $z=10$ according to Fig.~C1 of
D14, and from $n_{\rm DCBH}\sim 10^{-4}{\rm cMpc^{-3}}$ to $n_{\rm
DCBH}\sim 10^{-6}{\rm cMpc^{-6}}$ at $z=12$ according to Fig.~7 of A12.
Although the values of $n_{\rm DCBH}$ estimated according to D14 and A12 do not match
each other, both decrease by two or three orders of magnitude.  
The observed high-redshift SMBH number density is $n_{\rm
SMBH} \sim 10^{-9} {\rm cMpc^{-3}}$ at $z\sim 6$
(\cite{Fan:2001aa,Venemans:2013aa}), which is in the same order as
$n_{\rm DCBH}$ estimated according to D14.  In order to test the
scenario of SMBH formation via DCBH by comparing predicted $n_{\rm
DCBH}$ and observed $n_{\rm SMBH}$, it is crucial to more precisely
estimate $n_{\rm DCBH}$ in the light of our $J_{\rm LW,\,21}^{\rm
crit}=1400$.  We would like to note that the present-day SMBH number
density inferred from observed luminosity function of active galactic
nuclei is $n_{\rm SMBH}\sim 10^{-4}{\rm cMpc^{-3}}$
(\cite{Shankar:2009aa,Johnson:2013aa}), which is several orders of magnitude higher than 
our estimate of $n_{\rm DCBH}$, and it is thus unlikely that all
of the SMBHs are originated from DCBHs.

In order to achieve such strong external radiation as $J_{\rm
LW,\,21}\gtrsim1400$, source galaxies need to be very close to
the DCBH-forming halos.  In such cases, we expect that dynamical interactions
between the sources and clouds cannot be overlooked, and thus the above
extrapolation of the results of A12 and D14 may be no longer
correct. One possibility is large fraction of the pairs merge to single
larger halos due to the gravitational interactions before forming DCBHs,
as suggested by cosmological simulations (Chon et al. in prep.).
Another is that radiation from one of pairs of halos to
another realizes $J_{\rm LW,\,21}>1000$ during the synchronized
evolution of the pairs (\cite{Visbal:2014aa}).  In any case, it is
necessary to understand how strong external radiation $J_{\rm
LW,\,21}\gtrsim1400$ is realized, by studying further about the effects
of interactions between primordial-gas clouds and radiation sources.

In this paper, we have studied the dependence of $J^{\rm crit}$ on
spectra of external radiation.  However, $J^{\rm crit}$ also depends on
other physical conditions of the clouds and their environment.
\cite{Inayoshi:2011aa} found that $J^{\rm crit}$ increases in the
presence of cosmic-ray and/or X-ray, although it is not clear yet how
much cosmic-ray and/or X-ray are emitted from the same galaxy as the
source of radiation.  \cite{Omukai:2008aa} found the conditions on
metallicity allowed to form DCBHs, although it is not yet clear how
$J^{\rm crit}$ changes in the case that the metallicity is very small but not
exactly zero. D14 and \cite{Agarwal:2014aa} phenomenologically took into
account the effect of metal-enrichment from the same galaxy as the source
of radiation in their simulations.  S10 and \cite{Latif:2014ab} found
that $J^{\rm crit}$ has an order-of-magnitude scatter due to
three-dimensional structures of the clouds, such as shocks
and turbulence. R14 argued that self-shielding of the LW photons is
suppressed by the turbulence in the clouds due to the Doppler broadening of
lines.  On the other hand, the LW photons irradiated on the clouds may be
reduced by the Lyman series absorption of neutral hydrogen in the IGM.

In future studies, it is important to determine the probability
distribution of $J^{\rm crit}$, considering various physical conditions
of clouds and their environment.  By comparing $n_{\rm DCBH}$ predicted
with such probability distribution and observed $n_{\rm SMBH}$, 
high-precision test of the SMBH formation scenario via DCBH 
becomes possible.

\section*{Acknowledgments}

The authors would like to thank K. Toma for fruitful discussions and valuable
 comments.  This work is supported in part by the Grant-in-Aid from the
 Ministry of Education, Culture, Sports, Science and Technology (MEXT)
 of Japan (25287040 KO; 26287034 AKI).

\end{document}